\documentclass{emulateapj}

\newcommand{\myemail}{amber.straughn@asu.edu}
\shortauthors{Straughn et al.}
\shorttitle{Tadpole Galaxies in HUDF}

\newcommand{\etal}	{\mbox{et al.\,}}
\newcommand{\ie}	{\mbox{i.e.}}
\newcommand{\eg}	{\mbox{e.g.}}
\newcommand{\Mo}	{\mbox{M$_{\odot}$}}
\newcommand{\cle}	{\ensuremath{\lesssim}}

\newcommand{\fig}[1]{Figure~\ref{#1}}

\begin{document}

\title{Tracing Galaxy Assembly: Tadpole Galaxies in the Hubble Ultra-Deep Field}

\author{Amber N. Straughn, Seth H. Cohen, Russell E. Ryan Jr, Nimish P. Hathi,
Rogier A. Windhorst, \& Rolf A. Jansen} 

\affil{Department of Physics and Astronomy, Arizona State
University, Tempe, AZ 85281; \myemail}

\begin{abstract}
In the Hubble Ultra Deep Field (HUDF) an abundance of galaxies is seen
with  a  knot  at  one   end  plus  an  extended  tail,  resembling  a
tadpole. These  ``tadpole galaxies'' appear  dynamically unrelaxed ---
presumably  in an  early merging  stage ---  where  tidal interactions
likely  created  the  distorted  knot-plus-tail morphology.   Here  we
systematically select  tadpole galaxies from the HUDF  and study their
properties  as  a  function  of  their photometric  redshifts.   In  a
companion HUDF variability study  presented in this issue, Cohen \etal
(2005) revealed a total  of 45 variable objects believed  to be Active
Galactic Nuclei (AGN).  Here we show that this faint AGN sample has no
overlap  with  the  tadpole  galaxy  sample, as  predicted  by  recent
theoretical work.   The tadpole morphology --- combined  with the lack
of overlap with the variable  objects --- supports the idea that these
galaxies are in the process of an early-stage merger event, i.e., at a
stage that  likely precedes the  ``turn-on'' of any AGN  component and
the onset of any point-source  variability.  We show that the redshift
distribution  of tadpole galaxies  follows that  of the  general field
galaxy population, indicating that --- if most of the tadpole galaxies
are  indeed  dynamically young  ---  the  process  of galaxy  assembly
generally kept up with the  reservoir of available field galaxies as a
function of  cosmic epoch.  These new  observational results highlight
the importance  of merger-driven processes  throughout cosmic history,
and  are  consistent  with  a  variety of  theoretical  and  numerical
predictions.

\end{abstract}



\section{Introduction} \label{introduction}
The origin of disk galaxies has long been thought to occur through the
process of dissipational collapse in a Cold Dark Matter (CDM) universe
(White \& Rees, 1978).   Within this paradigm, hierarchical clustering
(Navarro, Frenk,  \& White 1997)  produces dark matter halos  in which
dissipational  collapse of  the  residual gas  occurs.  The  resulting
disks  retain the  kinematic  information of  their  host dark  matter
potential wells (Blumenthal  \etal 1986). Recent numerical simulations
have  resolved  some   long-standing  discrepancies  in  the  standard
dissipational  collapse  scenario  by  including  previously-neglected
energetic feedback from central supermassive black holes during galaxy
merging  events  (e.g. Robertson  \etal  2005).   In particular,  they
emphasize the relationship between the central black hole mass and the
stellar  velocity  dispersion, which  confirms  the  link between  the
growth of black holes and their host galaxies (di Matteo, Springel, \&
Hernquist  2005; Springel,  di  Matteo, \&  Hernquist 2005ab).   These
theoretical   predictions  place   merger-driven   scenarios  on   the
forefront, suggesting that galaxy merger activity is a crucial element
in  a cosmological  description of  the Universe.   The  present study
provides   observational  support  for   many  of   these  theoretical
predictions.

A large  abundance of galaxies in  the Hubble Ultra  Deep Field (HUDF;
Beckwith \etal 2005) appear dynamically unrelaxed, which suggests they
must  play  an  important  role  in  the  overall  picture  of  galaxy
evolution.    In  particular,   we   notice  many   galaxies  with   a
knot-plus-tail morphology.   This particular morphology  constitutes a
large, well-defined  subset of the  irregular and peculiar  objects in
the HUDF that  is uniquely measurable as described  in Section 3.  The
selection  of  this  specific  morphology  also ties  closely  to  the
numerical  simulations  described   above  (di  Matteo,  Springel,  \&
Hernquist  2005;  Springel, di  Matteo,  \&  Hernquist 2005ab),  which
predict  a  stage  of  merger-driven  galaxy  evolution  that  closely
resembles these tadpole galaxies in a distinct phase that does not yet
show AGN activity (as discussed further in Section 7).  In particular,
this  morphology appears  to represent  an \emph{early}  stage  in the
merging of  2 nearly-equal mass galaxies.   We systematically selected
galaxies displaying  this knot-plus-tail  morphology from the  HUDF; a
representative sample  of tadpoles is shown  in \fig{fig:tp1} (details
of sample selection are given in Section 3).  All the selected tadpole
galaxies contain the asymmetric,  pointlike source with a diffuse tail
morphology,  some with  multiple knots;  all of  which we  believe are
undergoing  recent interactions.  They  are mostly  linear structures,
some resembling the ``chain'' galaxies first reported by Cowie, Hu, \&
Songaila  (1995).  When  more  than two  clumps  come together,  these
objects may  be more  akin to the  luminous diffuse objects  and clump
clusters (Conselice \etal 2004;  Elmegreen, Elmegreen, \& Sheets 2004;
Elemegreen,  Elmegreen, \& Hirst  2004), or  other types  of irregular
objects  (van  den  Bergh  2002).   Elmegreen  \etal  (2005)  visually
classify 97 HUDF galaxies (down  to 10 pixels in size) as ``tadpoles''
and  126 as  ``double-clump.''   Some of  the  galaxies classified  by
Elmegreen \etal as ``double-clump'' were identified as tadpoles by our
code, due  either to the unresolved  nature of one  clump (which would
have been  detected as a ``tail''  in our analysis) or  to the diffuse
nature of  one end of the  object.  Since our goal  in selecting these
tadpoles  was to  sample galaxies  that had  \emph{recently} undergone
interaction,   inclusion  of  some   of  these   ``double-clumps''  is
warranted.  One high redshift object in our sample has been studied in
detail by Rhoads \etal (2005).   A few objects with multiple knots are
detected by  our selection  software, but the  majority have  a single
prominant knot with an extended tail.

In  this paper, we  present the  photometric redshift  distribution of
tadpole galaxies, and compare it with the redshift distribution of the
general  field   galaxy  population.   This  paper  is   organized  as
follows. In \S 2,  we describe the HUDF data, and in  \S 3 the tadpole
sample  selection. In \S  4, we  discuss why  the majority  of tadpole
galaxies are  likely \emph{not} chance alignments,  but instead mostly
dynamically  young  objects.  In  \S  5,  we  discuss  their  redshift
distribution, in \S  6 their relation to galaxy assembly,  and in \S 7
their possible relation to AGN growth.


\section{Hubble Ultra Deep Field Data} \label{data}

The Hubble  Ultra Deep  Field is  a 400 orbit  survey in  four filters
carried out  using the  Advanced Camera for  Surveys (ACS)  aboard the
\emph{Hubble Space Telescope} (\emph{HST})  of a single field centered
on    RA(J2000)=$03^{\rm   h}    32^{\rm    m}   39.\!\!^{\rm    s}0$,
Dec(J2000)=$-27^{\circ}  47' 29\farcs1$.   The 144-orbit  F775W ($i'$)
image is deepest,  followed by F850LP ($z'$; 144  orbits), F606W ($V$;
56 orbits), and F435W ($B$; 56 orbits). The HUDF reaches $\sim$1.0 mag
deeper  in $B$  and $V$  and  $\sim$1.5 mag  deeper in  $i'$ than  the
equivalent  filters   in  the   Hubble  Deep  Field   (Williams  \etal
1996). From  the $\sim$10,000 objects detected in  the HUDF (Koekemoer
2004), we will select the sample of tadpole galaxies and analyze their
properties using the $i'$-band  image, because it provides the highest
sensitivity of the four filters.  Yan \& Windhorst (2004b) discuss how
this results in a bias  against objects at z$\,\gtrsim5.5$.  This bias
is small,  and only  concerns the high  redshift tail of  the redshift
distribution. Note,  however, that tadpole  galaxies at z$\,\simeq5.5$
\emph{do} exist (\eg, Rhoads \etal 2005).


\section{Tadpole Sample Selection} \label{sampleselection}

The  first step  in  this  analysis is  to  systematically select  the
galaxies  that have  the  characteristic tadpole  shape.  We  selected
sources  in  the  F775W   ($i'$)  band  to  $i'_{AB}=28.0$  mag  using
\texttt{SExtractor} (Bertin \& Arnouts 1996).  The objects of interest
all have a bright ``knot'' at one end with an extended ``tail'' at the
other.  \texttt{SExtractor} selects  objects  from an  image based  on
different input parameters, and  adjusting them results in the desired
selection  of  sources.  The   crucial  parameter  at  this  stage  is
\textsf{\small  DEBLEND\_MINCONT}, which governs  the manner  in which
nearby peaks in  flux are considered part of a  single object and thus
are counted as one source. With the deblending parameter set to a high
value,  \texttt{SExtractor}  will  separate  nearby flux  maxima  into
separate sources.  In contrast,  when the deblending  is set to  a low
value,  the  program will  count  the  nearby  maxima largely  as  one
source. Two  different source catalogs are thus  generated: the highly
deblended catalog will contain  many point-like sources, including the
knots of potential tadpole  galaxies.  The catalog with low-deblending
will  contain  extended   sources,  including  the  tadpole  galaxies'
tails. The catalogs  contain many more sources than  the desired ones,
and  the correctly  shaped objects  must  be selected  from these  two
initial  catalogs.   The  desired  tadpole  galaxies   have  a  nearly
unresolved knot or concentration, and an extended tail, so these types
of  sources must  be selected  from the  initial catalog,  and related
spatially such that they represent  real objects. All of the following
procedures were performed using IDL.

Both input \texttt{SExtractor}  catalogs, described above, contain the
following information  about the selected  sources: $x$ and  $y$ pixel
locations, length of the semi-major  and semi-minor axis ($a$ and $b$)
of  \texttt{SExtractor}  ellipses, and  the  angle  ($\theta$) of  the
semi-major  axis   from  north  through  east.   The  following  input
parameters were  calibrated using a  training set of  tadpole galaxies
that were  manually selected by  visual inspection of portions  of the
HUDF.   First, the  knots of  the  tadpole galaxies  were selected  by
setting an  axis-ratio limit.  A ``knot'' was  defined to be  a source
from the  highly deblended  catalog with an  axis ratio  greater (\ie,
rounder) than  some critical value  (in our case, $b/a>0.70$).  In the
same way, the  ``tails'' needed to be elongated  objects, so a similar
procedure  was  performed  on   the  objects  from  the  catalog  with
low-deblending, but with the criterion that their $b/a< 0.43$. The two
new lists of correctly shaped  objects had to be related physically on
the image,  thus, a new  set of objects  was defined where a  knot was
within a certain  distance of the geometrical center  of a tail.  This
distance  was taken  to  be $<4a$  (in  semi-major axis  units of  the
tail). We  also required that  the knot be  at least $>0.1a$  from the
tail's  geometrical  center, since  we  are  searching for  asymmetric
objects, and  want to  eliminate upfront as  many of the  true edge-on
mid-type spiral disks as possible. The objects also must have the knot
near one end of the tail,  and this was accomplished by selecting only
those tails and knots that  had a relative angle $\theta$ --- measured
with  respect  to  the  semi-major  axis  of the  tail  ---  that  was
$\leq20^{\circ}$. This  step prevented including knots  and tails that
were close together on the image,  but not physically part of the same
galaxy.

\begin{table}
\caption{Galaxy Selector Input Parameters\tablenotemark{\dag}}
\label{table:param}
\begin{tabular*}{0.48\textwidth}
   {@{\extracolsep{\fill}}lr}
\hline
\hline
\multicolumn{1}{l}{Parameter} & \multicolumn{1}{r}{Value}\\
$ $&$ $\\
\hline
$b/a$ limit: knots & $>$0.70 \\
$b/a$ limit: tails & $<$0.43 \\
Distance to center (in a-axis units) & $<$4 \\
Angle difference $\theta$ (tail-knot) & $\leq20^{\circ}$ \\
Total number of tadpoles automatically selected & 154 \\
DEBLEND\_MINCONT (knots) & 0.000005 \\
DEBLEND\_MINCONT (tails) & 0.1 \\
\hline
\tablenotetext{\dag}{Table 1 lists \texttt{SExtractor} parameters used
in  the tadpole  sample  selection.   $b$ \&  $a$  are semi-major  and
semi-minor  axis lengths  of the  ellipses used  in  selecting tadpole
knots    and   tails.     DEBLEND\_MINCONT   (knots/tails)    is   the
\texttt{SExtractor} parameter that governs in what manner nearby peaks
in flux are considered part of a single object and are thus counted as
one source.}
\end{tabular*}
\end{table}

These  selection  criteria  provided  a  list  of  the  tadpole-shaped
galaxies.  The  final number  of  tadpoles  selected  depended on  the
selection program's input parameters for the limiting axis-ratios, the
distance at which the knots and tails were considered related, and the
angle  difference between  the knot  and  the semi-major  axis of  the
tail. The values of these parameters  are given in Table 1. With these
values,  the tadpole  galaxy  selection program  detected 154  sources
total. This sample was then refined as follows.

\begin{figure*}
\centerline{\epsfxsize=\hsize{\epsfbox{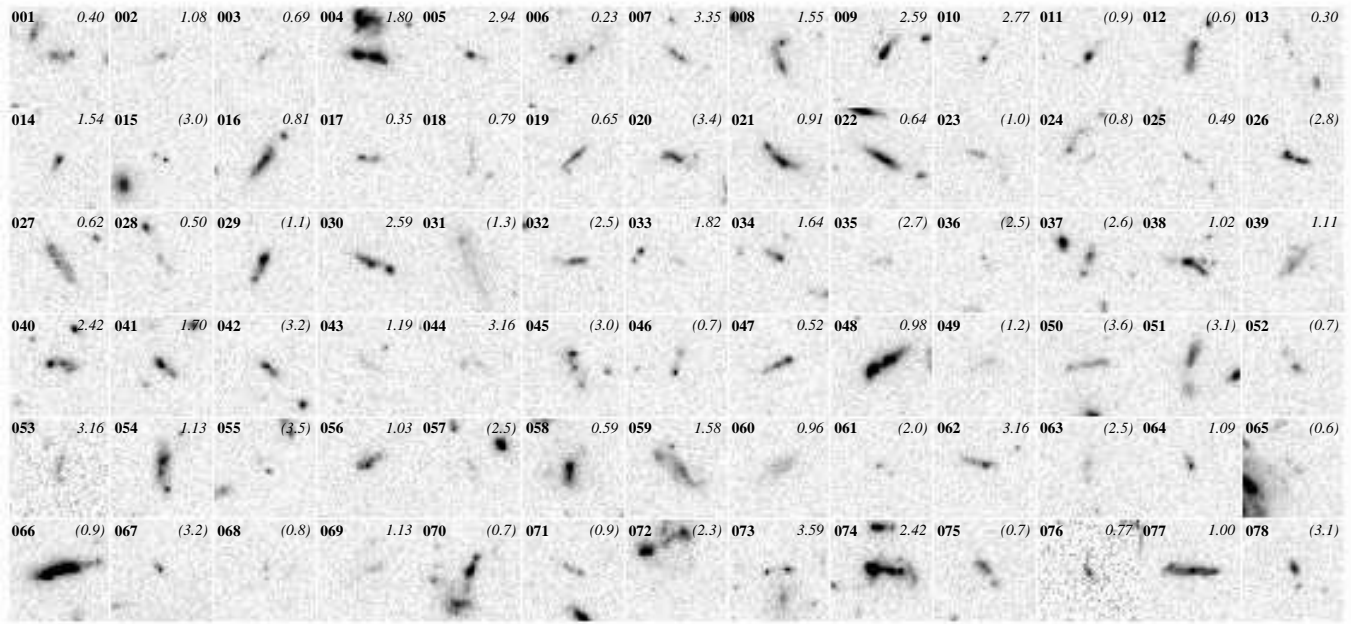}}}
\caption{F775W ($i'$)  band mosaic of  a subset of the  tadpole galaxy
sample in the HUDF.  Stamps  retain the orientation of the HUDF; north
is toward the top of the page  and east is to the left.  Index numbers
are  displayed in  the upper-left  corner of  the  stamps; photometric
redshifts  are  given in  upper-right  corners.  Parentheses  indicate
photometric redshifts  with errors $>$1 based  on HyperZ calculations.
Stamps are 3 arcsec on a  side.  A table of coordinates for the entire
tadpole  sample  is given  in  Section 3;  the  entire  sample of  165
tadpoles appears in color in the online supplement.  The vast majority
of  our   tadpole  sample  contains   the  distinctive  knot-plus-tail
morphology,  while sample  contamination  by normal  (non-interacting)
edge-on disk galaxies is less than 10\%.  }\label{fig:tp1}
\end{figure*}

A large majority of the 154 tadpole galaxy candidates selected had the
characteristic  elongated  knot-plus-tail  morphology, although  there
were some  anomalies.  In total, 14 (9\%)  obvious mis-detections were
visually rejected  because they  were very faint,  on the edge  of the
image, or  in the  outskirts of large  face-on spiral  galaxies, where
both  knotty  and  diffuse  regions  are common  and  spatially  close
together.  A  visual examination  of the field  also produced  25 more
tadpole  galaxies  not  found  by  the selector  program  due  to  the
inability  of  \texttt{SExtractor}  to correctly  separate  particular
point-like  sources  within  these  galaxies.   These  extra  selected
tadpole objects  visually obeyed the morphological  criteria that were
used to define  the main sample. Our total  final sample thus contains
165 tadpole galaxies, a subset of which is shown in \fig{fig:tp1}.  In
our final  sample, less than 10\%  of the selected  tadpoles appear as
normal  edge-on  disk galaxies;  the  vast  majority  have the  highly
asymmetric morphology.   In terms of visual  vs.\ automatic selection,
we find our sample to be  about 91\% (140/154) reliable and about 86\%
(154/179) complete.  The  final set of 165 tadpoles  galaxies will now
be studied  as a separate  class of dynamically unrelaxed  objects and
compared to the general field galaxy population in the HUDF.

\section{Why Tadpole Galaxies are Not Chance Alignments}
\label{whyyoung}

\begin{figure} 
\epsscale{1.2}
\plotone{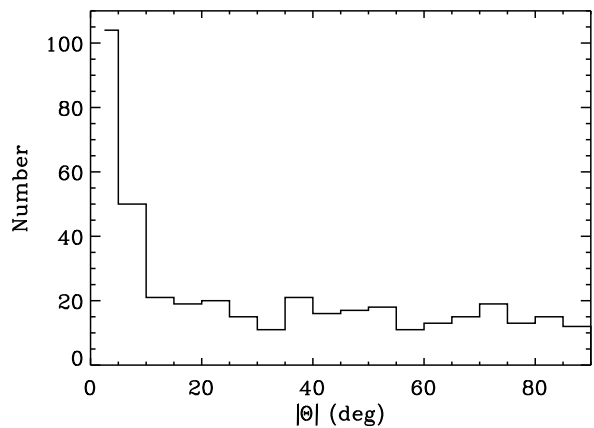}
\caption{Distribution of (the absolute  value of) angles ($\theta$) of
all off-centered knots  found within a radius $r\leq  4a$ ($\leq 2''$)
from  the center  of each  elongated  diffuse structure  in the  HUDF,
showing  a  clear  excess  of knots  near  $|\theta|\simeq  0^\circ$.}
\label{fig:theta}
\end{figure}

In this section we demonstrate  that these tadpole galaxies are likely
not chance alignnments of tails  and unrelated knots.  We first select
\emph{all} elongated  diffuse structures (``tails'') in  the HUDF, and
then  measure the  angle  $\theta$ of  the  nearest off-centered  knot
within  a  radius  $r\leq  4a$  ($\leq 2''$).   Chance  alignments  of
unrelated tails and knots would  show a random distribution of angles;
however,  \fig{fig:theta} shows  that there  clearly is  an  excess of
knots  near $|\theta|\simeq  0^\circ$.  The  excess peak  contains 154
knots, while  the average number of knots  with $|\theta|\ge 10^\circ$
is $\sim$15  per $5^\circ  \theta$-bin.  \fig{fig:theta} thus  shows a
significant overabundance  of knots near the end  of elongated diffuse
structures  as compared  to randomly  distributed knots.   Hence, this
physically  meaningful result  suggests that  the majority  of tadpole
galaxies are  not just chance alignments of  unrelated knots. Instead,
we  believe they  are mostly  linear structures  which  are undergoing
interactions.  When  compared to models of galaxy  mergers (di Matteo,
Springel, \&  Hernquist 2005; Springel, di Matteo,  \& Hernquist 2005;
Robertson  \etal 2005,  Hopkins  \etal 2005),  these objects  strongly
resemble dynamically young objects in the early stages of merging.


\section{The Redshift Distribution of Tadpole Galaxies} 
\label{redshiftdistribution}

To  investigate  the occurrence  of  tadpole  galaxies throughout  the
history  of the universe,  we calculate  photometric redshifts  of all
HUDF galaxies  to $i'_{AB}=28.0$ mag.  All  photometric redshifts were
calculated  from  the   HUDF  $BViz$(+$JH$)  photometry  using  HyperZ
(Bolzonella \etal  2000). In order to  investigate associated redshift
errors,   we   compared  our   photometric   redshifts  to   published
spectroscopic redshifts for CDFS 70 objects. We find an rms scatter of
0.15    for     the    fractional    photometric     redshift    error
$\delta$=(photoz-specz)/(1+specz) if all  70 objects are included, and
0.10 when we  reject a few of the most  obvious outliers.  This result
is fully consistent with prior claims of photometric redshift accuracy
in  the literature (Lanzetta  \etal 1997,  Mobasher \etal  2004).  The
accuracy of our photometric redshift estimates depends on the accuracy
of the measured magnitudes in  each of the available filters.  It also
is largely independent  of the shape of an  object (although magnitude
errors for more extended, lower  surface brightness objects tend to be
somewhat  larger  than those  for  more  concentrated, higher  surface
brightness objects of the same total magnitude).

\begin{figure}
\epsscale{1.2}
\plotone{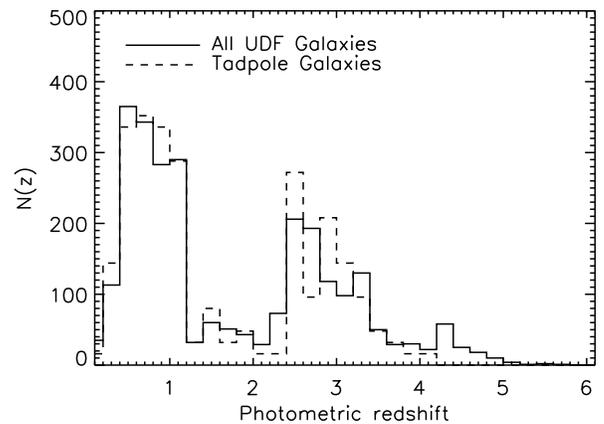}
\caption{Photometric   redshift  distribution   of  galaxies   in  the
HUDF. The solid black histogram shows the redshift distribution of all
HUDF  field galaxies  to  $i'_{AB}=28.0$, while  the dashed  histogram
shows the  redshift distribution of  the tadpole galaxies.  The latter
was multiplied by 16$\times$ for best comparison of its shape with the
redshift distribution of the field galaxies.}\label{fig:hist} 
\end{figure}

\begin{figure} 
\epsscale{1.2}
\plotone{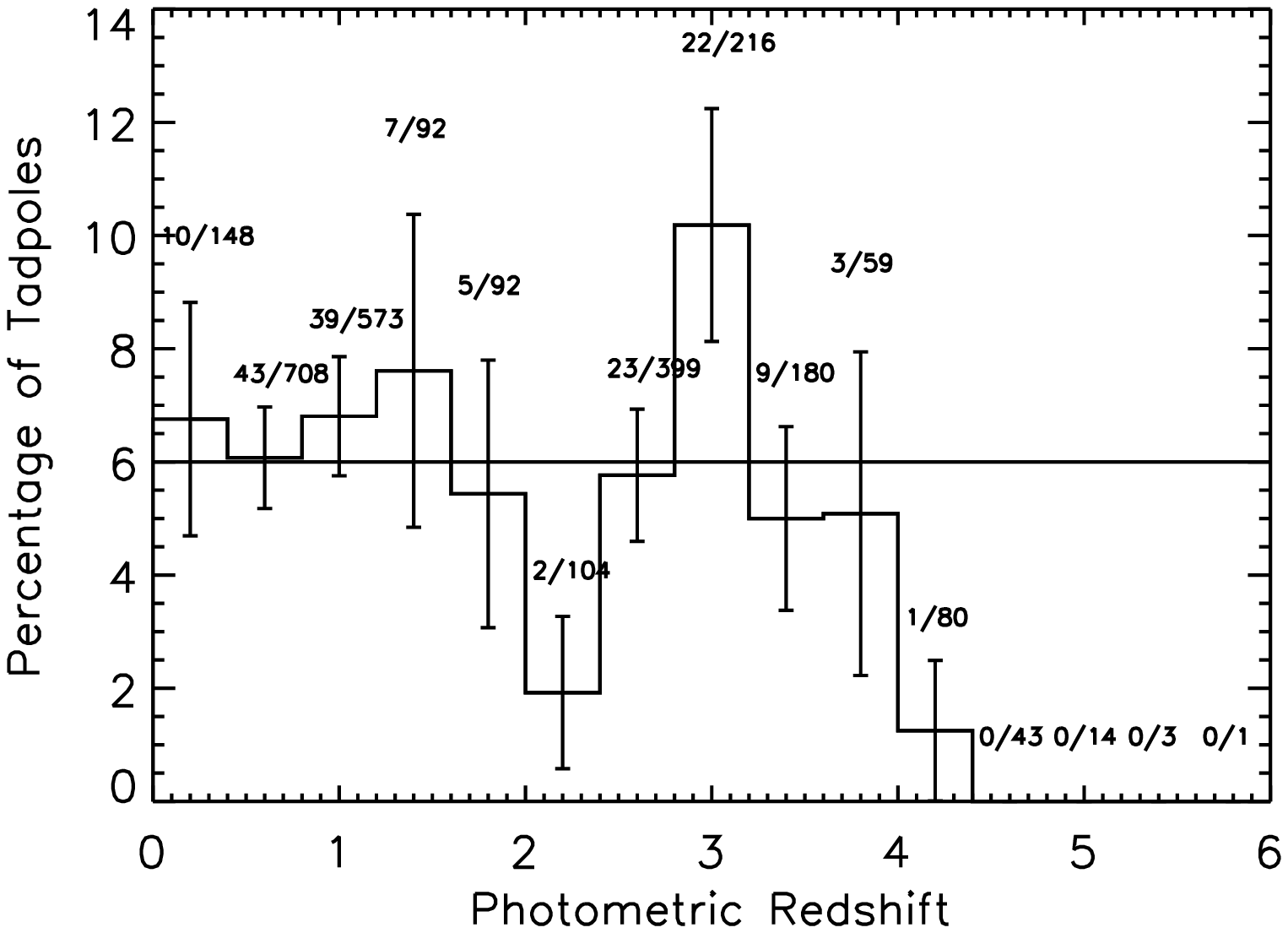} 
\caption{Percentage of total galaxies  that are tadpoles is plotted as
a function of photometric  redshift. Within the statistical errors, on
average  about  6\%  of all  galaxies  are  seen  as tadpoles  at  all
redshifts.}\label{fig:per}
\end{figure}

The redshift distribution  of all galaxies in the  HUDF (solid line in
\fig{fig:hist})   is   as  expected,   with   the   primary  peak   at
$0.5\leq\,$z$\,\leq1.0$   and   a    generally   declining   tail   at
z$\,\simeq4$--5.   These trends  were  also seen  in  the general  HDF
redshift  distribution of  faint field  galaxies (Driver  \etal 1998).
Also  apparent  is  a  lack  of  objects  at  z$\,\simeq1$--2  due  to
unavailable   UV   spectral   features  crossing   the   $BViz$(+$JH$)
filters. This occurs because the  HUDF does not have deep enough F300W
or  $U$-band (ultra-violet)  data,  unlike the  situation  in the  HDF
(Williams \etal 1996). This redshift bias, however, is the \emph{same}
for  both the  tadpole and  the general  field galaxy  populations. In
\fig{fig:hist},  the tadpole  galaxy distribution  is multiplied  by a
factor  of 16 for  best comparison  with that  of the  field galaxies.
Within the  available statistics,  the redshift distribution  shape of
the tadpole galaxies follows that  of the general field galaxies quite
closely. This suggests that if tadpole galaxies are indeed dynamically
young objects  related to early-stage  mergers, they may occur  in the
same   proportion   to   the    field   galaxy   population   at   all
redshifts.  Tadpole galaxies  may  therefore be  good  tracers of  the
galaxy assembly process.  The  ratio of the two redshift distributions
N(z) was calculated  as well, and the resulting  percentage of tadpole
galaxies  is  plotted  in  \fig{fig:per}  as a  function  of  redshift
together  with  the statistical  errors.  Overall,  the percentage  of
tadpole  galaxies is roughly  constant at  $\sim$6\% with  redshift to
within the  statistical errors  for the redshift  range probed  in our
study ($0.1\leq$z$\leq4.5$).

\section{Tadpole Galaxies as Tracers of Galaxy Assembly} \label{tracers}

The fact that about 6\% of  all field galaxies are seen in the tadpole
stage is  a measurement  with potentially important  consequences.  In
light  of  simulations  by   Springel  \etal  (2005)  that  predict  a
tadpole-like  stage $\simeq$0.7 Gyr  after a  major merger  begins, we
suggest  that   this  particular  tadpole   morphology  represents  an
early-merger stage of two galaxies  with comparable mass.  If this 6\%
indicates  the fraction of  time that  an average  galaxy in  the HUDF
spends in  an early-merger stage during  its lifetime, and  if most of
these low-luminosity  objects started forming the bulk  of their stars
at  the end of  the reionization  epoch at  z$\simeq6-7$ (e.g.  Yan \&
Windhorst 2004a,  2004b), then  each galaxy would  spend about  6\% of
12.9 Gyr (i.e. 0.8 Gyr)  since z$\simeq7$ in a distinctly recognizable
merger or tadpole stage. At  the median redshift at which the tadpoles
are seen  (z$_{med}$ $\simeq1.6$; see \fig{fig:hist}),  each object is
then seen at an age of about 4 Gyr if born at z$\simeq$7. Each tadpole
is $\simeq1$" (or $\simeq8$ kpc) across (\fig{fig:tp1}), and given the
fluxes  measured,  each clump  in  a  tadpole  has roughly  M$\,\simeq
10^8$--$10^9\,\Mo$  in  stars  (see,  e.g.,  Papovich,  Dickinson,  \&
Ferguson 2001  who estimated stellar masses of  lyman break galaxies).
For these  rough estimates of their physical  parameters, the freefall
timescale   for  each  tadpole   is  roughly   $\tau\cle$  (few$\times
10^7$)--$10^8$ years,  or $\simeq$6\% of  the galaxy lifetime  at that
redshift.  Hence, if  every  galaxy is  seen  in a  tadpole stage  for
$\simeq$0.8  Gyr   of  its  lifetime,  then  it   may  have  undergone
$\sim$10-30 mergers  during its lifetime.   During the early  stage of
each merger, it  would be temporarily seen as  a tadpole. More complex
mergers may  lead to  irregular/peculiar and train-wreck  type objects
and the  luminous diffuse objects  or clump clusters, which  are among
the  type  of  objects  that  dominate  the  galaxy  counts  at  faint
magnitudes (Driver  \etal 1998).  In  this paper, we limit  the sample
selection to two clumps passing  by each other, which we believe leads
to  the more  uniquely  classifiable tadpole  morphology.  Given  that
tadpoles only trace a certain  type and stage of merging galaxies, the
above  statistics  are likely  a  lower limit  on  the  number of  all
mergers.  In  summary, each  galaxy seen today  may have had  of order
one- to two- dozen mergers since  most of its Population II stars were
born  at z$\simeq$7,  and  given  the small  masses  and short  merger
timescales involved, would then be  seen as tadpole galaxies for about
6\% of their life-time.

\fig{fig:per} suggests that tadpole galaxies --- if indeed dynamically
young objects --- appear to occur  in the same proportion to the field
galaxy  population at  all redshifts  probed in  this  study.  Tadpole
galaxies  may therefore  be  good  tracers of  the  process of  galaxy
assembly.  This  implies that  the process of  galaxy assembly  --- as
traced  by  tadpole  galaxies  ---  keeps up  with  the  reservoir  of
available    galaxies    as     a    function    of    redshift    for
$0.1\!\leq\!z\!\leq\!4.5$.   Our result  is  in excellent  agreement
with  the  predictions  of  Robertson  \etal (2005)  that  describe  a
merger-driven scenario  to build up  disk galaxies, and  is consistent
with Rhoads \etal (2005) who conclude that their z=5.4 galaxy (tadpole
\# 165 in Table 2) is strongly indicative of a galaxy in assembly.

\section{Relation between Tadpole Galaxies and Active Galactic Nuclei}

In a companion paper in this issue, Cohen \etal (2005) present a study
of  the  variable  objects in  the  HUDF  which  have a  point  source
component with  a measurably variable  flux on timescales  of 0.4--3.5
months, which roughly corresponds  to 0.5--5.5 weeks in the rest-frame
at  the median redshift  of the  sample (z$_{med}$  $\simeq1.5-2$). In
particular,  they  found  45  plausibly variable  objects  among  4644
galaxies  to $i'_{AB}=28.0$ mag.   They argue  that these  objects are
most likely variable  because they host weak AGN.  Sometimes these AGN
are  in  the galaxy  center,  but often  they  occur  off-center in  a
dynamically unrelaxed system. This prompts the question: What fraction
of tadpole galaxies contains a variable weak AGN in its knots? This is
a critical  issue, because it is  widely believed that  the process of
merging in galaxies  can also disturb the inner  accretion disk around
the supermassive black  hole (SMBH) and switch on  the AGN.  Among our
165 tadpole  galaxies, none  coincide with the  sample of  45 variable
objects  or with the  x-ray sources  in the  Chandra Deep  Field South
(Alexander, D.M. \etal  2005).  Recent state-of-the-art hydrodynamical
models (di  Matteo, Springel, \& Hernquist 2005;  Springel, di Matteo,
\&  Hernquist 2005; Hopkins  \etal 2005)  suggest that  during (major)
mergers, the  black hole accretion  rate peaks considerably  after the
merger  started,   and  after  the  star  formation   rate  (SFR)  has
peaked. Specifically, their models suggest that, for massive galaxies,
a  tadpole stage  is seen  typically about  0.7 Gyr  after  the merger
started, but $\sim$0.9 Gyr before  the SMBH accretes most of its mass,
which  is when  the galaxy  displays  strong AGN  activity. Since  the
lifetimes   of   QSO's   and    radio-galaxies   are   known   to   be
$\cle\,$(few$\times  10^7$)--$10^8$ years  (Martini, P.  2004, Grazian
\etal 2004), these hydrodynamical models thus imply that the AGN stage
is expected to occur  considerably \emph{after} (i.e., $\ge$1 Gyr) the
early-merger  event during  which the  galaxy is  seen in  the tadpole
stage.  The observed lack of  overlap between the tadpole galaxies and
the AGN  sample in the  HUDF provide direct observational  support for
this prediction.  Recent studies by Grogin \etal (2005) find asymmetry
(A)  values that  are similar  between AGN  and non-AGN  samples; this
result is consistent with our study in light of the theoretical models
mentioned above,  which indicate that  AGN activity is seen  only well
\emph{after} the  merger has  taken place and  the galaxy  has settled
into  a more dynamically  relaxed state.   In addition,  Hopkins \etal
(2005) have  quantified the  timescales that  quasars will  be visible
during  merging  events, noting  that  for  a  large fraction  of  the
accretion time, the quasar  is heavily obscured.  In particular, their
simulations show that  during an early merging phase  --- our observed
tadpole  phase  --- the  intrinsic  quasar  luminosity  peaks, but  is
completely  obscured.  Only  after  feedback from  the central  quasar
clears out  the gas will  the object become  visible as an  AGN.  This
should  be   observable  by  the   Spitzer  Space  Telescope   in  the
mid-infrared (IR) as a  correspondingly larger fraction of IR-selected
obscured faint QSO's.

In conclusion,  tadpole galaxies  are a class  of easily-identifiable,
dynamically  young objects that  exist throughout  the history  of the
Universe and are good tracers of galaxy assembly.  They provide strong
observational   support   for  the   validity   of  recent   numerical
simulations, and highlight the importance of mergers to the process of
galaxy assembly and AGN growth.

\section{Acknowledgements}\label{Acknowledgements}

We thank the STScI staff, and in particular Steve Beckwith, Ray Lucas,
and Massimo  Stiavelli, for their persistent efforts  to implement the
Hubble Ultra  Deep Field in the  best possible way.  This research was
partially  funded  by  NASA  grants  GO-9793.08-A  and  AR-10298.01-A,
awarded by  STScI, which is operated  by AURA for  NASA under contract
NAS 5-26555;  as well as by the  NASA Space Grant program  at ASU, and
the Harriet G. Jenkins Predoctoral Fellowship Program.

\begin{table}
\caption{Global Properties of Tadpole Galaxies}
\label{table2}
\begin{tabular*}{0.98\textwidth}
   {@{\extracolsep{\fill}}ccccc|ccccc|ccccc}
\hline
\hline
\multicolumn{1}{c}{ID}&\multicolumn{1}{c}{RA}&\multicolumn{1}{c}{Dec}&\multicolumn{1}{c}{$i'_{AB}$}&\multicolumn{1}{c|}{$z$\tablenotemark{\ddag}}  & \multicolumn{1}{c}{ID}&\multicolumn{1}{c}{RA}&\multicolumn{1}{c}{Dec}&\multicolumn{1}{c}{$i'_{AB}$}&\multicolumn{1}{c|}{$z$\tablenotemark{\ddag}} & \multicolumn{1}{c}{ID}&\multicolumn{1}{c}{RA}&\multicolumn{1}{c}{Dec}&\multicolumn{1}{c}{$i'_{AB}$}&\multicolumn{1}{c}{$z$\tablenotemark{\ddag}}\\
\multicolumn{1}{c}{$ $}&\multicolumn{1}{c}{($h:m:s$)}&\multicolumn{1}{c}{($\degr$ '  '')}&\multicolumn{1}{c}{(mag)}&\multicolumn{1}{c|}{$ $} & \multicolumn{1}{c}{$ $}&\multicolumn{1}{c}{($h:m:s$)}&\multicolumn{1}{c}{($\degr$ '  '')}&\multicolumn{1}{c}{(mag)}&\multicolumn{1}{c|}{$ $} & \multicolumn{1}{c}{$ $}&\multicolumn{1}{c}{($h:m:s$)}&\multicolumn{1}{c}{($\degr$ '  '')}&\multicolumn{1}{c}{(mag)}&\multicolumn{1}{c}{$ $}\\
$ $&$ $&$ $&$ $&$ $&$ $&$ $&$ $&$ $&$ $&$ $ & $ $ & $ $& $ $& $ $\\
\hline
1 & 03:32:30.118 & -27:47:17.61 & 24.86 & 0.40 & 	56 & 03:32:36.680 & -27:45:39.20 & (26.0) & 1.03 & 	111 & 03:32:40.820 & -27:49:04.40 & (24.4) & 0.98 \\
2 & 03:32:30.162 & -27:47:36.22 & 27.28 & 1.08 & 	57 & 03:32:36.683 & -27:47:38.53 & 27.25 & 2.51 & 	112 & 03:32:40.920 & -27:48:23.90 & (25.2) & 1.11 \\
3 & 03:32:30.266 & -27:47:50.45 & 27.45 & 0.69 & 	58 & 03:32:36.860 & -27:46:04.00 & (26.0) & 0.59 & 	113 & 03:32:40.929 & -27:46:33.76 & 26.90 & 1.17 \\
4 & 03:32:30.674 & -27:47:42.30 & 24.14 & 1.80 & 	59 & 03:32:36.920 & -27:46:34.79 & 24.85 & 1.58 & 	114 & 03:32:41.000 & -27:45:44.10 & 26.69 & 3.77 \\
5 & 03:32:30.995 & -27:48:03.88 & 26.38 & 2.94 & 	60 & 03:32:37.138 & -27:46:25.94 & 26.01 & 0.96 & 	115 & 03:32:41.118 & -27:47:34.59 & 24.17 & 0.73 \\
6 & 03:32:31.108 & -27:47:58.64 & 25.01 & 0.23 & 	61 & 03:32:37.240 & -27:48:54.80 & 27.43 & 2.03 & 	116 & 03:32:41.126 & -27:45:58.71 & 27.28 & 1.29 \\
7 & 03:32:31.190 & -27:48:01.19 & 26.33 & 3.35 & 	62 & 03:32:37.340 & -27:45:49.80 & (26.5) & 3.16 & 	117 & 03:32:41.354 & -27:48:49.53 & 26.04 & 0.32 \\
8 & 03:32:31.399 & -27:47:13.45 & 25.35 & 1.55 & 	63 & 03:32:37.347 & -27:47:39.45 & 26.89 & 2.51 & 	118 & 03:32:41.374 & -27:47:38.12 & 25.45 & 2.94 \\
9 & 03:32:31.530 & -27:47:58.40 & (25.5) & 2.59 & 	64 & 03:32:37.350 & -27:45:37.90 & (26.7) & 1.09 & 	119 & 03:32:41.480 & -27:46:42.40 & (26.9) & 1.19 \\
10 & 03:32:31.853 & -27:47:42.06 & 26.78 & 2.77 & 	65 & 03:32:37.352 & -27:48:38.22 & 26.92 & 0.61 & 	120 & 03:32:41.487 & -27:45:56.28 & 27.58 & 0.78 \\
11 & 03:32:31.883 & -27:47:39.00 & 25.93 & 0.87 & 	66 & 03:32:37.409 & -27:47:41.65 & 23.52 & 0.90 & 	121 & 03:32:41.507 & -27:46:53.52 & 27.59 & 0.60 \\
12 & 03:32:32.125 & -27:47:27.94 & 25.29 & 0.62 & 	67 & 03:32:37.460 & -27:47:23.30 & (26.8) & 3.18 & 	122 & 03:32:41.560 & -27:49:23.35 & 25.19 & 3.37 \\
13 & 03:32:32.218 & -27:46:50.67 & 26.23 & 0.30 & 	68 & 03:32:37.546 & -27:46:36.98 & 27.85 & 0.78 & 	123 & 03:32:41.583 & -27:46:39.94 & 24.86 & 0.86 \\
14 & 03:32:32.500 & -27:47:02.00 & (26.5) & 1.54 & 	69 & 03:32:37.570 & -27:49:11.50 & 26.38 & 1.13 & 	124 & 03:32:41.595 & -27:49:01.80 & 24.94 & 0.91 \\
15 & 03:32:32.601 & -27:47:11.24 & 27.43 & 2.96 & 	70 & 03:32:37.591 & -27:47:39.49 & 24.22 & 0.67 & 	125 & 03:32:41.596 & -27:48:49.85 & 26.55 & 0.39 \\
16 & 03:32:32.704 & -27:48:14.77 & 24.85 & 0.81 & 	71 & 03:32:37.735 & -27:48:30.27 & 26.71 & 0.93 & 	126 & 03:32:41.598 & -27:48:08.09 & 25.48 & 0.99 \\
17 & 03:32:32.739 & -27:46:40.70 & 26.34 & 0.35 & 	72 & 03:32:37.813 & -27:47:57.34 & 26.85 & 2.29 & 	127 & 03:32:41.724 & -27:46:56.50 & 26.53 & 1.19 \\
18 & 03:32:32.959 & -27:47:02.08 & 27.12 & 0.79 & 	73 & 03:32:37.832 & -27:45:52.97 & 25.77 & 3.59 & 	128 & 03:32:41.762 & -27:47:27.67 & 25.51 & 2.85 \\
19 & 03:32:33.004 & -27:48:18.71 & 25.97 & 0.65 & 	74 & 03:32:37.881 & -27:48:53.11 & 23.59 & 2.42 & 	129 & 03:32:41.791 & -27:47:38.69 & 26.28 & 2.41 \\
20 & 03:32:33.067 & -27:47:43.96 & 25.50 & 3.37 & 	75 & 03:32:37.949 & -27:47:33.16 & 25.79 & 0.70 & 	130 & 03:32:41.805 & -27:47:23.88 & 27.03 & 2.94 \\
21 & 03:32:33.086 & -27:48:13.01 & 24.79 & 0.91 & 	76 & 03:32:38.020 & -27:45:09.30 & 26.13 & 0.77 & 	131 & 03:32:41.960 & -27:45:48.82 & 26.94 & 3.59 \\
22 & 03:32:33.112 & -27:48:23.05 & 24.82 & 0.64 & 	77 & 03:32:38.096 & -27:45:26.83 & 24.61 & 1.00 & 	132 & 03:32:42.476 & -27:47:44.63 & 25.79 & 3.28 \\
23 & 03:32:33.212 & -27:47:11.07 & 26.51 & 0.97 & 	78 & 03:32:38.312 & -27:47:28.11 & 26.05 & 3.14 & 	133 & 03:32:42.510 & -27:47:03.10 & (26.2) & 2.49 \\
24 & 03:32:33.228 & -27:47:25.27 & 27.97 & 0.82 & 	79 & 03:32:38.376 & -27:49:15.24 & 27.19 & 0.46 & 	134 & 03:32:42.788 & -27:48:56.89 & 25.66 & 2.96 \\
25 & 03:32:33.541 & -27:46:40.55 & 27.64 & 0.49 & 	80 & 03:32:38.430 & -27:46:34.80 & (24.4) & 2.58 & 	135 & 03:32:42.910 & -27:47:01.77 & 26.92 & 2.94 \\
26 & 03:32:33.706 & -27:47:56.64 & 25.49 & 2.77 & 	81 & 03:32:38.541 & -27:46:16.10 & 27.26 & 0.76 & 	136 & 03:32:42.930 & -27:48:19.22 & 26.82 & 1.53 \\
27 & 03:32:33.911 & -27:46:17.05 & 25.28 & 0.62 & 	82 & 03:32:38.559 & -27:47:30.25 & 24.66 & 2.96 & 	137 & 03:32:43.086 & -27:46:46.12 & 25.91 & 0.60 \\
28 & 03:32:34.047 & -27:46:42.73 & 26.39 & 0.50 & 	83 & 03:32:38.608 & -27:48:04.05 & 26.11 & 3.37 & 	138 & 03:32:43.108 & -27:46:14.10 & 26.05 & 3.18 \\
29 & 03:32:34.180 & -27:48:03.20 & (25.7) & 1.11 & 	84 & 03:32:38.659 & -27:49:18.86 & 23.78 & 0.61 & 	139 & 03:32:43.302 & -27:46:43.46 & 27.08 & 0.55 \\
30 & 03:32:34.295 & -27:46:47.67 & 25.05 & 2.59 & 	85 & 03:32:38.816 & -27:45:24.50 & 27.08 & 1.06 & 	140 & 03:32:43.395 & -27:47:14.41 & 23.79 & 0.95 \\
31 & 03:32:34.438 & -27:46:59.48 & 25.11 & 1.33 & 	86 & 03:32:38.930 & -27:48:56.80 & (25.9) & 2.81 & 	141 & 03:32:43.948 & -27:47:13.69 & 24.34 & 0.48 \\
32 & 03:32:34.673 & -27:47:25.27 & 26.16 & 2.51 & 	87 & 03:32:39.194 & -27:48:54.93 & 27.57 & 0.53 & 	142 & 03:32:43.953 & -27:46:45.38 & 27.82 & 0.41 \\
33 & 03:32:34.704 & -27:47:59.83 & 27.80 & 1.82 & 	88 & 03:32:39.233 & -27:48:49.83 & 25.53 & 2.94 & 	143 & 03:32:43.985 & -27:46:33.06 & 23.24 & 0.06 \\
34 & 03:32:34.790 & -27:47:24.30 & (26.5) & 1.64 & 	89 & 03:32:39.325 & -27:45:55.16 & 24.85 & 0.57 & 	144 & 03:32:44.560 & -27:46:23.53 & 25.82 & 0.58 \\
35 & 03:32:34.909 & -27:48:06.77 & 27.53 & 2.74 & 	90 & 03:32:39.350 & -27:45:55.40 & (26.4) & 0.57 & 	145 & 03:32:44.645 & -27:47:02.36 & 25.67 & 2.61 \\
36 & 03:32:34.981 & -27:47:03.03 & 27.78 & 2.51 & 	91 & 03:32:39.404 & -27:49:06.49 & 25.06 & 0.98 & 	146 & 03:32:44.772 & -27:47:08.89 & 26.04 & 0.73 \\
37 & 03:32:35.253 & -27:47:14.14 & 26.00 & 2.58 & 	92 & 03:32:39.405 & -27:46:22.41 & 25.69 & 3.28 & 	147 & 03:32:44.910 & -27:47:58.10 & (27.2) & 1.89 \\
38 & 03:32:35.260 & -27:46:54.30 & (25.7) & 1.02 & 	93 & 03:32:39.485 & -27:47:34.63 & 25.90 & 0.90 & 	148 & 03:32:44.999 & -27:46:29.53 & 25.82 & 0.23 \\
39 & 03:32:35.280 & -27:48:57.25 & 25.84 & 1.11 & 	94 & 03:32:39.490 & -27:49:23.24 & 26.32 & 0.50 & 	149 & 03:32:45.237 & -27:46:39.19 & 26.43 & 3.14 \\
40 & 03:32:35.353 & -27:48:54.56 & 25.29 & 2.42 & 	95 & 03:32:39.530 & -27:47:39.70 & (25.9) & 0.52 & 	150 & 03:32:45.246 & -27:46:43.93 & 25.61 & 0.40 \\
41 & 03:32:35.520 & -27:47:53.80 & (25.9) & 1.70 & 	96 & 03:32:39.533 & -27:49:31.24 & 25.69 & 0.72 & 	151 & 03:32:45.919 & -27:47:30.18 & 26.02 & 2.44 \\
42 & 03:32:35.670 & -27:46:47.70 & (26.2) & 3.16 & 	97 & 03:32:39.540 & -27:46:04.90 & (26.3) & 0.70 & 	152 & 03:32:45.945 & -27:47:20.42 & 24.88 & 2.58 \\
43 & 03:32:35.878 & -27:49:01.58 & 27.48 & 1.19 & 	98 & 03:32:39.580 & -27:49:12.83 & 25.44 & 1.07 & 	153 & 03:32:45.975 & -27:46:57.60 & 23.58 & 1.43 \\
44 & 03:32:35.881 & -27:45:57.00 & 27.17 & 3.16 & 	99 & 03:32:39.600 & -27:45:54.60 & (24.3) & 0.39 & 	154 & 03:32:46.016 & -27:47:06.38 & 25.70 & 2.77 \\
45 & 03:32:35.988 & -27:47:25.53 & 25.38 & 2.96 & 	100 & 03:32:39.656 & -27:45:29.97 & 25.29 & 0.35 & 	155 & 03:32:46.103 & -27:47:08.05 & 27.03 & 1.16 \\
46 & 03:32:36.169 & -27:48:17.30 & 26.53 & 0.73 & 	101 & 03:32:39.723 & -27:45:46.98 & 24.86 & 0.92 & 	156 & 03:32:46.384 & -27:48:11.19 & 25.68 & 0.41 \\
47 & 03:32:36.193 & -27:46:08.88 & 25.89 & 0.52 & 	102 & 03:32:39.775 & -27:46:18.16 & 27.65 & 3.86 & 	157 & 03:32:46.482 & -27:47:44.45 & 26.16 & 0.50 \\
48 & 03:32:36.267 & -27:48:34.18 & 23.86 & 0.98 & 	103 & 03:32:39.829 & -27:45:31.74 & 25.89 & 0.90 & 	158 & 03:32:47.247 & -27:47:57.83 & 25.23 & 0.90 \\
49 & 03:32:36.272 & -27:47:09.55 & 27.13 & 1.19 & 	104 & 03:32:39.909 & -27:46:56.06 & 27.81 & 2.88 & 	159 & 03:32:47.386 & -27:47:26.02 & 25.34 & 4.06 \\
50 & 03:32:36.290 & -27:47:53.48 & 26.11 & 3.63 & 	105 & 03:32:39.920 & -27:48:58.90 & (26.4) & 0.90 & 	160 & 03:32:48.340 & -27:47:28.44 & 26.64 & 0.79 \\
51 & 03:32:36.301 & -27:47:22.40 & 25.19 & 3.10 & 	106 & 03:32:40.200 & -27:46:02.90 & (26.1) & 0.98 & 	161 & 03:32:37.734 & -27:47:06.96 & 23.33 & 0.60 \\
52 & 03:32:36.462 & -27:48:32.06 & 26.56 & 0.67 & 	107 & 03:32:40.391 & -27:48:29.47 & 25.18 & 2.59 & 	162 & 03:32:41.865 & -27:46:51.10 & 23.52 & 0.71 \\
53 & 03:32:36.567 & -27:49:17.54 & 26.35 & 3.16 & 	108 & 03:32:40.562 & -27:46:28.56 & 27.25 & 2.51 & 	163 & 03:32:42.993 & -27:47:09.73 & 23.78 & 2.74 \\
54 & 03:32:36.613 & -27:48:01.42 & 24.79 & 1.13 & 	109 & 03:32:40.670 & -27:46:41.49 & 26.48 & 2.51 & 	164 & 03:32:41.077 & -27:48:52.98 & 20.58 & 0.28 \\
55 & 03:32:36.661 & -27:48:03.11 & 27.28 & 3.54 & 	110 & 03:32:40.761 & -27:48:36.62 & 25.74 & 2.90 & 	165 & 03:32:33.257 & -27:47:24.69 & (27.3) & 5.4\tablenotemark{\dag} \\
\hline
\tablenotetext{}{NOTE: parentheses indicate estimated aperture magnitudes for visually selected objects}
\tablenotetext{\dag}{redshift from Rhoads \etal 2005}
\tablenotetext{\ddag}{Photometric redshifts computed from HyperZ (Bolzonella \etal  2000)}
\end{tabular*}
\end{table}

\end{document}